\documentclass[preprint,sort&compress]{elsarticle}

\usepackage{amsmath}

\journal{Nucl. Instrum. and Meth. A.}

\begin{document}

\begin{frontmatter}

\title{Test of a single module of the J-PET scanner based on plastic scintillators}

\cortext[cor1]{Corresponding author.}

\author[WFAIS]{P.~Moskal}
\author[WFAIS]{Sz.~Nied\'zwiecki}
\author[WFAIS]{T.~Bednarski}
\author[WFAIS]{E.~Czerwi\'nski\corref{cor1}}
\ead{eryk.czerwinski@uj.edu.pl}
\author[WFAIS,PAN]{\L .~Kap\l on}
\author[WFAIS]{E.~Kubicz}
\author[WFAIS]{I.~Moskal}
\author[WFAIS]{M.~Pawlik-Nied\'zwiecka}
\author[WFAIS]{N.G.~Sharma}
\author[WFAIS]{M.~Silarski}
\author[WFAIS]{M.~Zieli\'nski}
\author[WFAIS]{N.~Zo\'n}
\author[WFAIS]{P.~Bia\l as}
\author[WFAIS]{A.~Gajos}
\author[WCHUJ]{A.~Kochanowski}
\author[WFAIS]{G.~Korcyl} 
\author[WFAIS]{J.~Kowal}
\author[SWIERK]{P.~Kowalski}
\author[WFAIS]{T.~Kozik}
\author[WFAIS]{W.~Krzemie\'n}
\author[WCHUJ]{M.~Molenda}
\author[WFAIS]{M.~Pa\l ka}
\author[SWIERK]{L.~Raczy\'nski}
\author[WFAIS]{Z.~Rudy}
\author[WFAIS]{P.~Salabura}
\author[WFAIS]{A.~S\l omski} 
\author[WFAIS]{J.~Smyrski}
\author[WFAIS]{A.~Strzelecki}
\author[WFAIS,PAN]{A.~Wieczorek}
\author[SWIERK]{W.~Wi\'slicki}

\address[WFAIS]{Faculty of Physics, Astronomy and Applied Computer Science,
        Jagiellonian University, 30-059 Cracow, Poland}
\address[PAN]{Institute of Metallurgy and Materials Science of Polish Academy of Sciences, Cracow, Poland.}
\address[WCHUJ]{Faculty of Chemistry, Jagiellonian University, 30-060 Cracow, Poland}
\address[SWIERK]{\'Swierk Computing Centre, National Centre for Nuclear Research, 05-400 Otwock-\'Swierk, Poland}

\begin{abstract}
Time of Flight 
Positron Emission Tomography scanner 
based on plastic scintillators is being developed at the Jagiellonian University
by the J-PET collaboration.
The main challenge of the conducted research 
lies in the elaboration of a method allowing application 
of plastic scintillators for the detection of low energy gamma quanta. 
In this paper we report on tests of a single detection 
module built out from the BC-420 plastic scintillator strip
(with dimensions of 5~x~19~x~300~mm$^3$) 
read out at two ends by Hamamatsu R5320 photomultipliers. 
The measurements were performed using collimated beam of annihilation quanta from the $^{68}$Ge isotope and applying 
the Serial Data Analyzer (Lecroy SDA6000A) which enabled sampling of signals with 50 ps intervals. 
The time resolution of the prototype module
was established to be better than 80~ps ($\sigma$) 
for a single level discrimination. The spatial resolution of the determination of the hit position along the strip was determined 
to be about 0.93~cm~($\sigma$) for the annihilation quanta. 
The fractional energy resolution for the energy E deposited by the annihilation quanta via 
the Compton scattering amounts to 
$\sigma(E)/E \approx 0.044 / \sqrt{E(MeV)}$ and corresponds to the $\sigma(E)/E$ of 7.5\% at the Compton edge.
\end{abstract}

\begin{keyword}
\texttt{scintillator detectors} \sep \texttt{J-PET} \sep \texttt{Positron Emission Tomography}
\end{keyword}

\end{frontmatter}

\section{Introduction}
\begin{figure}[h!]
\includegraphics[width=0.45\textwidth]{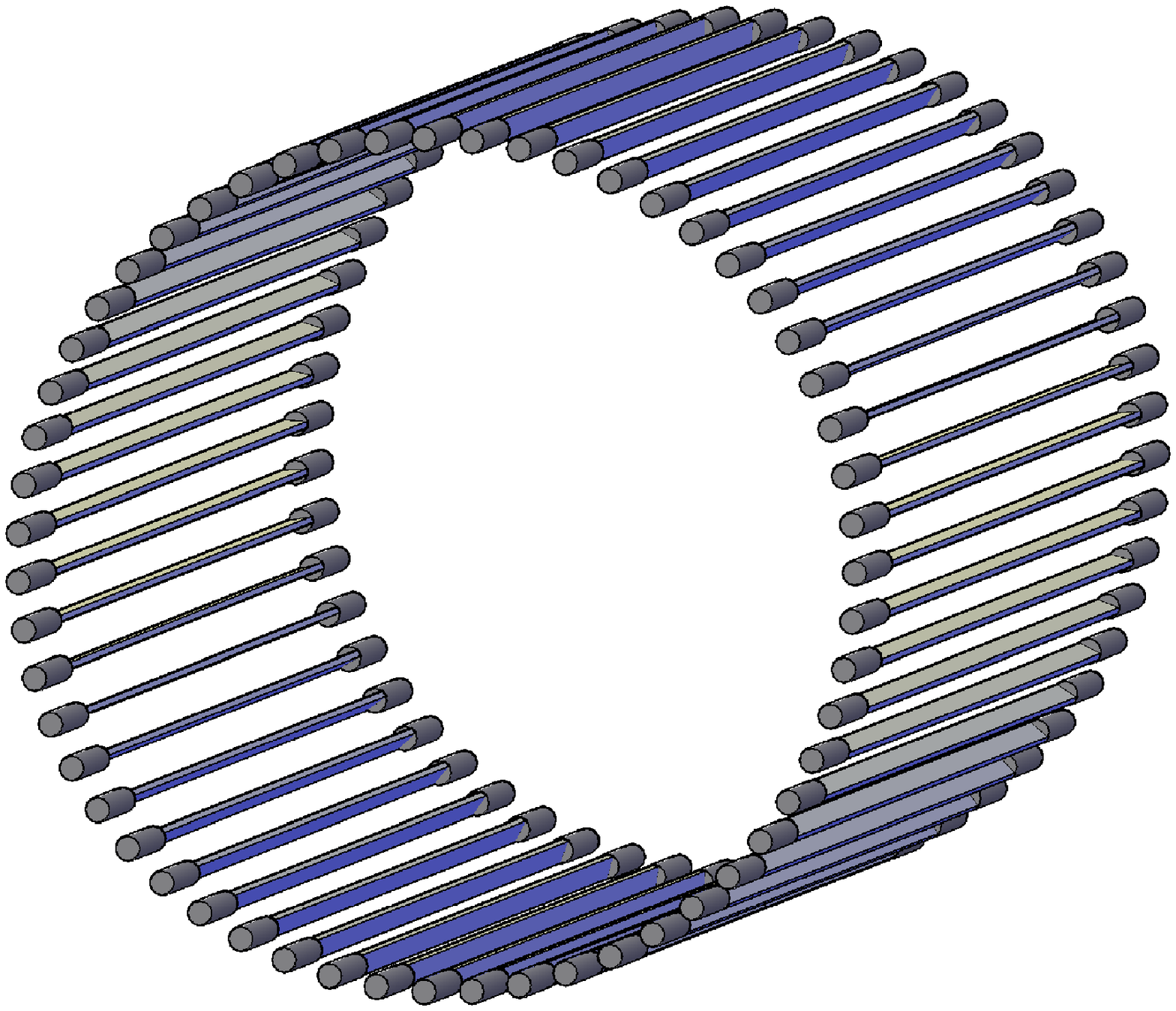} \hspace{-0.55cm} \includegraphics[width=0.58\textwidth]{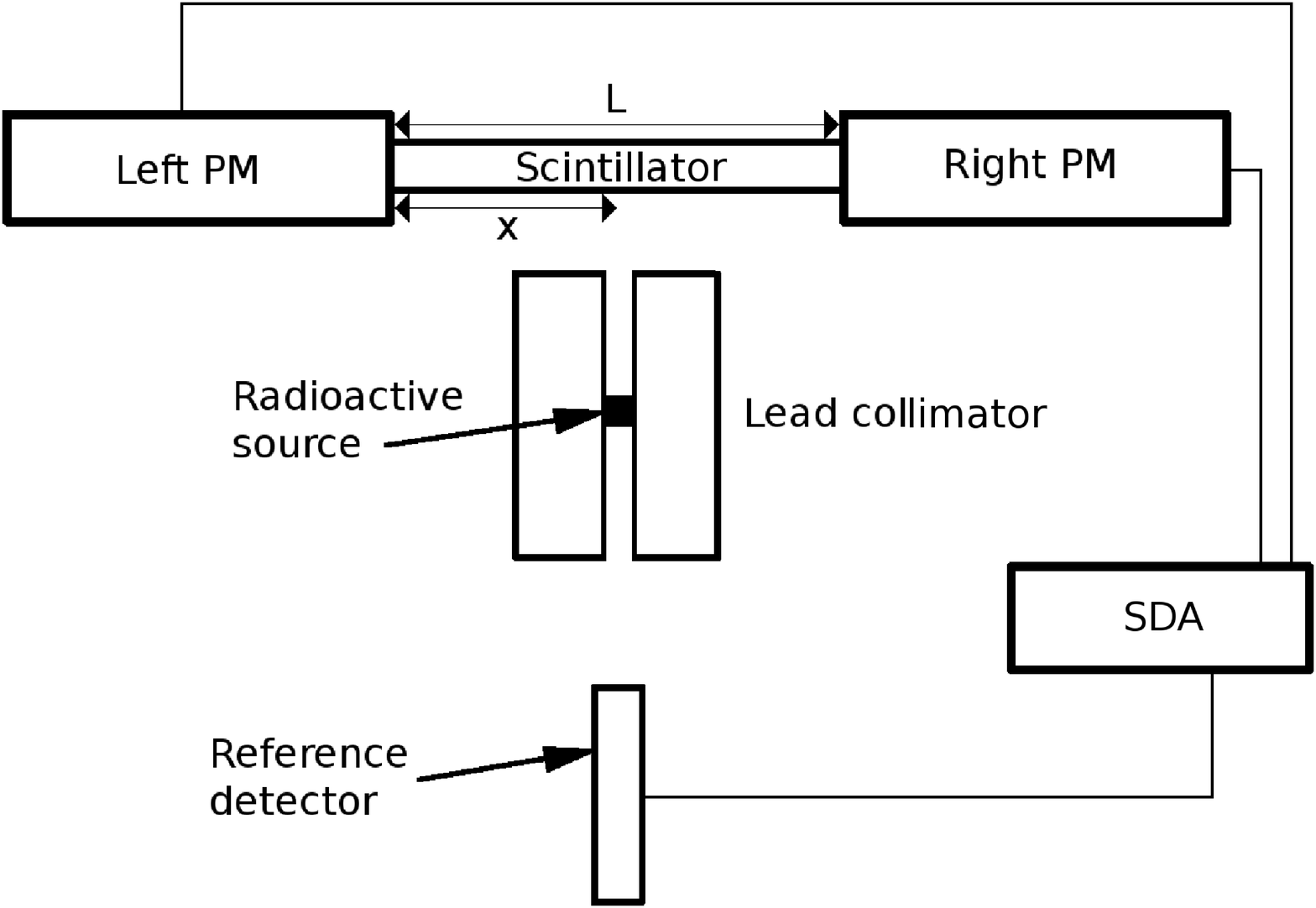}\\
\caption{(Left) Schematic view of the single layer of the J-PET scanner. 
(Right) General scheme of the experimental setup used to test performance 
of a single detection module. Radioactive source is held in the lead collimator. 
\textit{Abbreviations:} PM and SDA denote photomultiplier and Serial Data Analyzer (Lecroy SDA6000A), respectively.
}
\label{uklad}
\end{figure}

Positron Emission Tomography is currently one of the best suited medical examination 
methods for tumor detection. 
Nowadays commercial PET scanners are made of crystal scintillators arranged in 
a ring surrounding the patient~\cite{Conti2009,Humm2003,Karp2008,Townsend2004}.
New generation of PET scanners for the image reconstruction not only utilizes information about the hit position 
of gamma quanta in the detectors but also takes advantage of the measurement of the time differences (TOF) 
between the interactions of annihilation quanta in the detectors~\cite{Conti2011}. 
This improves image reconstruction by increasing the signal to background ratio~\cite{Conti2009,Karp2008,Moses2003}.  
A typical TOF resolution of presently used TOF-PET detectors amounts to about 500~ps~(FHWM)~\cite{Conti2011},
and there is a continuous endeavor to improve it (see e.g. results for small size 
crystals~\cite{ContiErikson2009,Schaart2010,Schaart2009,Moses2008,Kuhn2006,MoszynskiSzczesniak2011}).
The Jagiellonian-PET (J-PET) collaboration aims at the construction of the TOF-PET 
scanner with a large field of view (up to about 1 m) 
and a superior TOF resolution by application of fast plastic scintillators instead of 
organic crystals. The detector will be built out from strips of plastic scintillators forming 
a diagnostic chamber~\cite{PCT2010,JPET-Genewa,NovelDetectorSystems,StripPETconcept,TOFPETDetector}
as shown schematically in the left panel of Fig.~\ref{uklad}. 

A single detector module of the J-PET detector consists of a plastic scintillator strip 
read out by photomultipliers at both ends. Plastic scintillators are less efficient for the detection of gamma quanta 
than crystals but they posses better timing properties and allow to build large acceptance detectors in a cost efficient way.
Therefore, a PET scanner based on plastic scintillators constitutes a promising solution in view of the 
TOF resolution and construction of the scanner allowing for simultaneous imaging of the whole human body.
Development of a cost-effective whole body PET scanner  is
a technological challenge and there are various non-standard techniques being tested such as detectors based on straw tubes drift chambers~\cite{Lacy2001,Shehad2005} 
or large area resistive plate chambers~\cite{Belli2006,Blanco2009}. 

The J-PET detector with plastic scintillators arranged axially as shown in Fig.~\ref{uklad}
possesses also another advantage. Its diagnostic chamber is free of any electronic devices and magnetic materials, 
thus giving unique possibility for simultaneous imaging of PET and MRI as well as PET and CT
in a way different from so far developed
configurations~\cite{Crosetto2003,Gilbert2006,Judenhofer2008,Marsden2002,Pichler2008,Quick2011,Townsend2008}.
A similar axial arrangement with crystal scintillators of the length of 10~cm
is being developed by the AX-PET collaboration aiming at improvement of resolution and sensitivity~\cite{AX-PET}. 

Detectors based on plastic scintillators are commonly used in nuclear and particle physics experiments, 
however, due to negligible probability of photo-electric effect, 
their potential  for registration of low energy gamma quanta (in the range of fraction of MeV)
was so far not explored except for few publications concentrated
on the light  propagation studies~e.g.~\cite{NIM2008Moszynski}, 
or callibration methods~\cite{NIM2008,NIM2011,NIMKudomi}. 
In this paper we show that plastic scintillators can be used for building  
large area detection systems with very good time, position and energy resolution for the registration of low energy gamma quanta.

In Section 2 a comprehensive description of experimental setup used for investigations is presented. 
Next, sections 3, 4 and 5 include description of methods and results for the determination of the energy, time and position resolution, respectively.

\section{Experimental setup}
\begin{figure}[h!]
	\centerline{\includegraphics[width=0.7\textwidth]{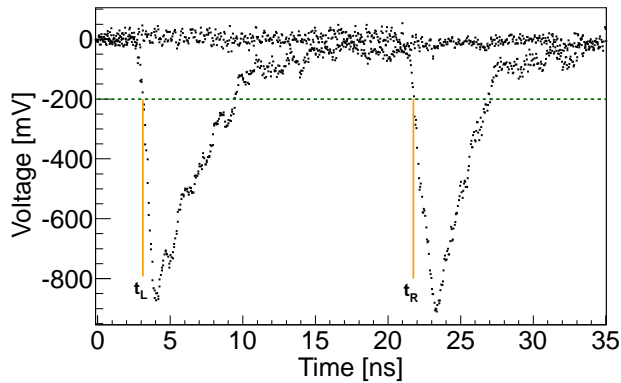}}
\caption{
Example of signals acquired from left and right photomultiplier 
when irradiating the center of the scintillator strip. 
$t_L$ and $t_R$ denote times at which left and right signal, respectively, 
cross the reference voltage indicated by dashed green horizontal line. 
For better visibility the signals were separated from each other by 19~ns.
(For interpretation of the references to color in this figure caption, the reader is referred to the web version of this paper.)
\label{shapes}
}
\end{figure}

General scheme of the experimental setup used for tests of a single module is presented in the right panel of Fig.~\ref{uklad}.  
A prototype module consists of a BC-420~\cite{SaintGobain} scintillator strip with dimensions of 
5~mm~x~19~mm~x~300~mm and of two Hamamatsu photomultipliers R5320~\cite{Hamamatsu} connected optically 
to the most distant ends of the scintillator strip via optical gel EJ-550. 
The module is tested with annihilation quanta from the $^{68}$Ge source 
placed inside a lead collimator which can be moved along the scintillator 
by means of a dedicated mechanical construction. A collimated beam emerging through 1.5~mm wide and 20~cm 
long slit is used for irradiating desired points across the strip. A coincident registration 
of signals from the tested module and a reference detector allows for a suppression of signals from other than annihilation quanta 
to the negligible level. The reference detector consists of a scintillator strip with 
thickness of 4~mm connected via light guide to the photomultiplier. The reference detector is fixed 
to the collimator by means of an aluminum arm allowing to keep the relative setting between the collimator and 
the reference detector unchanged while moving the collimator along the tested scintillator strip. 
In this way the same collimating properties are ensured at every position of irradiation. 
Signals from photomultipliers are probed with 50~ps intervals by means of Serial Data Analyzer 
(LeCroy SDA6000A).
Exemplary sampled signals from the middle of the scintillator are shown in Fig.~\ref{shapes}. 
For the full J-PET detector a dedicated electronics~\cite{palka,korcyl}
and analysis framework~\cite{krzemien,czerwinski} 
for data collecting and processing is being developed.

In the J-PET detector a hit position along the scintillator strip, 
as well as an annihilation point along the line-of-response, will be reconstructed based on the measurement 
of times of light signals arrival to photomultipliers.  Therefore, the detector needs to be optimized 
for the best timing properties. 
Moreover, for building a device with a large field of view, 
a weak light attenuation in the scintillator material is mandatory.
These requirements led us to a choice 
of the BC-420~\cite{SaintGobain}
(equivalent of EJ-230~\cite{Eljen})
plastic scintillator 
as the most suitable among the currently available ones. 
The rise time and bulk attenuation length of light signals in this scintillator 
amount to 0.5~ns and 110~cm~\cite{SaintGobain}, respectively. 
For the light detection, the Hamamatsu 
R5320 
photomultipliers~\cite{Hamamatsu} were chosen
with the rise time and the transit time spread of 0.7~ns and 0.16~ns, respectively~\cite{Hamamatsu}. 
The rise time of signals shown in Fig.~\ref{shapes} is equal to about 1~ns, 
as expected from 0.5 ns rise time of a light pulse in the BC-420 scintillator 
convoluted with 0.7 ns rise time of signals from the Hamamatsu R5320 photomultiplier.
The decay time for the BC-420 scintillator amounts to 1.5~ns~\cite{SaintGobain}.
In a good approximation~\cite{Moszynski} an observed signal is a convolution 
of the  Gaussian and exponential functions and of the single photoelectron response of the photomultiplier.
The single photoelectron signals were measured using a method described in Ref.~\cite{CalibrationPMT},
and values of a rise time of 0.7~ns and FWHM of 1.5~ns were observed
in agreement with the values given in catalog~\cite{Hamamatsu}. 
Moreover, we have checked that the observed signals shown in Fig.~\ref{shapes}
are consistent with the expectation for the decay time of 1.5~ns. 
The studies presented in this paper were conducted for the scintillator strip wrapped with the tyvek foil.
For further details about 
properties of the used photomultipliers and scintillators in view of the construction 
of the J-PET detector the interested reader is referred to~\cite{CalibrationPMT,lukasz}. 

\section{Energy resolution}
Energy resolution depends predominantly on the number of photoelectrons released from photocathodes of both photomultipliers. 
The larger this number the better is the energy resolution due to decrease of the statistical fluctuation of the number of signal carriers. 
Therefore, for the consideration of the energy resolution it is natural to express the energy deposition 
in terms of the number of photoelectrons and to use an arithmetic mean as a measure of the deposited energy: 
\begin{equation}
	E_{deposited} = \alpha  \frac{(N_L + N_R)}{2},
\end{equation}
where $\alpha$, $N_L$ and $N_R$  
denote an energy calibration factor, and 
number of photoelectrons registered 
at left and right side of the scintillator, respectively.   
For scintillator detectors the Fano factor is equal to one and therefore,
in case of uncorrelated errors of $N_L$ and $N_R$ the fractional energy resolution would read:
\begin{equation}
	\frac{\sigma(E_{deposited})}{E_{deposited}} = \frac{1}{\sqrt{N_L+N_R}} = \frac{\sqrt{\alpha/2}}{\sqrt{E_{deposited}}}.
        \label{beta1}
\end{equation}
Consequently, the energy resolution as a function of the deposited energy may be approximately parametrized as:
\begin{equation}
	\frac {\sigma (E_{deposited}) }{E_{deposited}} = \frac {\beta }{\sqrt{E_{deposited}}}, 
	\label{beta}
\end{equation}
where $\beta$ is an effective coefficient which in general may differ from $\sqrt {\alpha/2}$.  
\begin{figure}[h!]
\centerline{\includegraphics[width=0.6\textwidth]{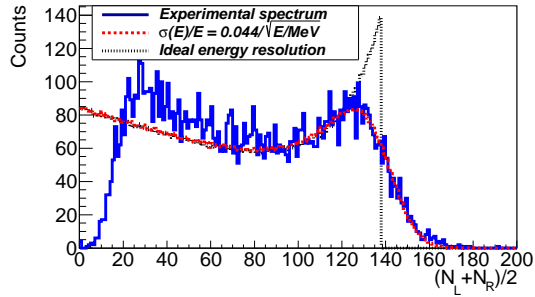}}
\caption{Distribution of arithmetic mean of the number of photoelectrons produced at photocathodes 
of left and right photomultipliers. The spectrum was obtained by irradiating the center of the scintillator strip with the collimated 
beam of annihilation quanta. 
As indicated in the legend, solid and dashed lines indicate experimental and simulated spectra, respectively.
More detailed explanation is given in the text. (For interpretation of the references to color in this figure caption, the reader is referred to the web version of this paper.)
\label{GandA}}
\end{figure}

The value of $\beta$ for the tested detector
was obtained by comparing experimental distribution of $(N_L+N_R)/2$ (Fig.~\ref{GandA}) 
with the simulated histogram of deposited energy where $\alpha$, 
$\beta$, and normalization constant $A$ were treated as free parameters. 
A fit was conducted with a Neyman $\chi^2$ statistics defined as follows:     
\begin{equation}
	\chi^2(\alpha,\beta,A) = \sum_{i}\frac{(A \cdot  N_{sim}(i\cdot\alpha,\beta) - N_{exp}(i))^2}{N_{exp}(i)},
\end{equation}
where $i$ denotes the $i^{th}$ bin of the histogram $N_{exp}$. 
The simulated distribution of energy deposition of the annihilation quanta $N_{sim}(E,\beta)$ 
was obtained based on the Klein-Nishina formula~\cite{Klein} convoluted with the detector resolution 
parametrized by Eq. \ref{beta}. Due to the large number of generated events 
the statistical uncertainties of simulated distributions are negligible. 
The fit was performed in the range from 90 to 150 photoelectrons. 
The lower range of the spectrum was not taken into account since it is enhanced 
by events with the scattering very close to the scintillator surface and
by signals originating from gamma quanta scattered 
in the collimator and in material surrounding the detector.  The best fit was obtained for $\beta = 0.044$. 
Dashed and dotted lines in Fig.~\ref{GandA} show simulated energy loss spectra for the ideal detector and the detector 
with the fractional energy resolution of $\sigma (E)/E = 0.044 / \sqrt{E(MeV)}$, as obtained from the fit.  

It is also worth to stress that the determined value of $\beta$
is fairly close to the result expected for the fully uncorrelated 
errors of $N_L$ and $N_R$ which (compare Eq.~\ref{beta1} and~\ref{beta}) 
gives 
$\beta = \sqrt{\alpha/2} \approx \sqrt{0.341/138/2} \approx 0.035$,  
where $\alpha$  
is estimated knowing that $E_{deposited}$~=~0.341~MeV at the Compton edge 
and that the corresponding mean value of photoelectrons 
from left and right photomultipliers amounts to about 138 (see Fig.~\ref{GandA}). 
The number of photoelectrons for each measured signal was determined based on the known average charge of signals 
induced by single photons determined using the method described in Ref.~\cite{CalibrationPMT}.
To calculate the number of photoelectrons the charge of each  measured signal was divided by the average charge of the single photoelectron signal. 
In order to measure a charge spectrum originating from the single photoelectron
we have inserted between the tested photomultiplier and the scintillator an aperture with a hole with a diameter of 0.6~mm.
The ratio of the area of the hole (0.28~mm$^2$) and the side of the scintillator (95~mm$^2$) was equal to about 340,
thus providing that for the typical event out of about 300 photons reaching the edge of the scintillator
only zero, one or very rarely two photoelectrons were released from the photocathode.  
The experimental spectrum shown in Fig.~\ref{GandA} is suppressed in the range below 20 photoelectrons due to the triggering conditions.
The superimposed red dashed line indicates the distribution simulated based on the Klein-Nishina formula~\cite{Klein} 
convoluted with the detector response with a resolution of $\sigma(E)/E$ as indicated in the figure. The dotted line denotes the theoretical distribution 
of the energy of electrons scattered via the Compton effect 
by the gamma quantum with an energy of 511 keV.
The observed number of photoelectrons is consistent with 
rough estimations of about 149 photoelectrons at the Compton edge (0.341~MeV)
which can be derived 
taking into account that (i) the light output of BC-420 scintillator 
equals to about 10,000 photons per MeV~\cite{SaintGobain} for electrons,
(ii) the fraction of light which can be conducted via internal reflections to the edge
in the rectangular strip surrounded by air is equal to about $\sqrt{1 - (1/n)^2} - \frac{1}{2}$~=~0.27
(with refractive index n~=~1.58~\cite{SaintGobain}), (iii) 
the quantum effciency of Hamamatsu R5320 photomultipliers 
is equal to about 0.2~\cite{Hamamatsu} 
at  400 nm, 
and (iv) the bulk light attenuation length is equal to about 110~cm~\cite{SaintGobain}  
where on the average the light travels about 23~cm 
from the center to the edge ($e^{-0.21}$~=~0.81).

In case of the reconstruction of the tomographic image it was estimated that only signals with $E_{deposited} > 0.2$~MeV 
will be used in order to decrease the noise caused by the scattering 
of the annihilation quanta in the patient's body~\cite{TOFPETDetector}. 
In the energy range from 0.2~MeV to 0.34~MeV the value of $\beta~=~0.044$ 
gives fractional energy resolution of $\sigma(E)/E$ ranging from about 10\%  to 7.5\%, respectively. 
In the discussed case energy deposition of 0.2~MeV corresponds to about 81 photoelectrons.
Yet, in the further analysis, for conservative estimation of the time and position resolution we have selected 
signals with at least 75 photoelectrons.

\section{Time resolution}
\begin{figure}[b!]
	\centerline{\includegraphics[width=0.49\textwidth]{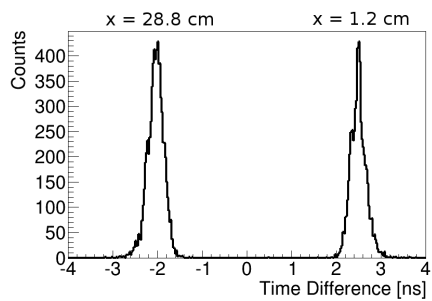}}
\caption{Distributions of time difference $\Delta t = (t_R - t_L)$ 
for two positions as indicated in the figure and discribed in the text.
\label{twoOnOne}
}
\end{figure}

\begin{figure}[t!]
\begin{tabular}{c c}
	\includegraphics[width=0.49\textwidth]{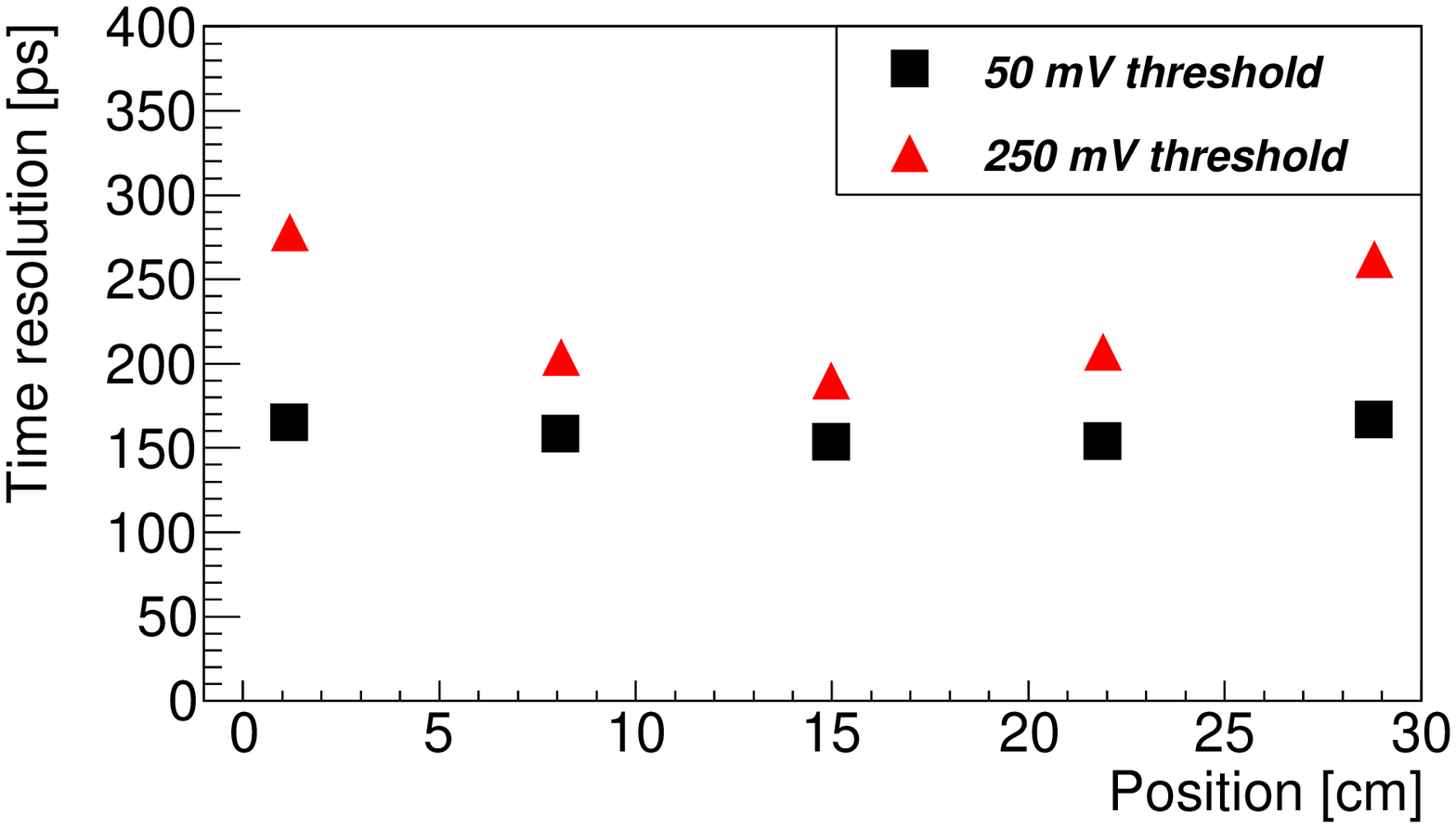} & \includegraphics[width=0.49\textwidth]{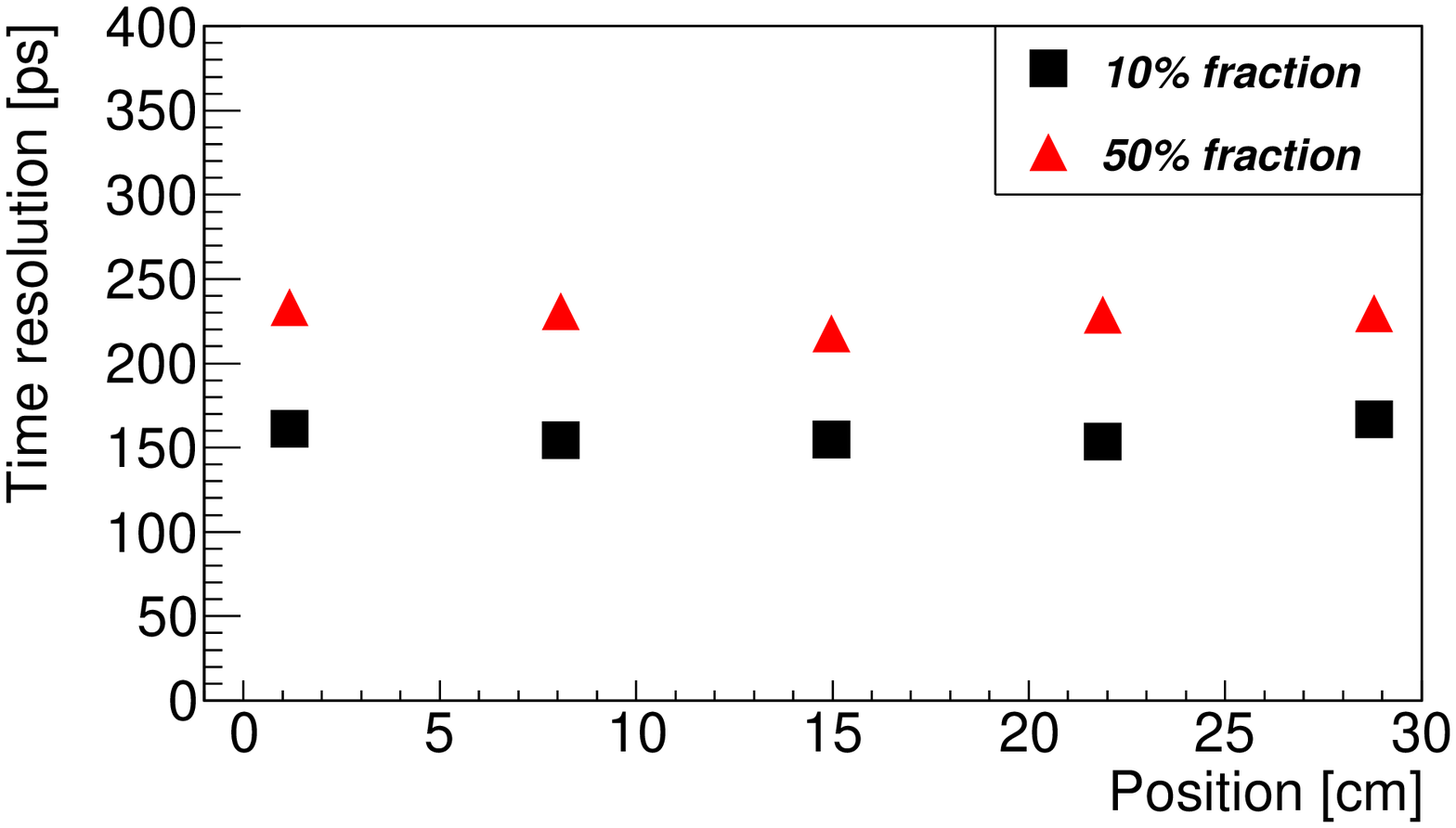}\\
\end{tabular}
\caption{ Left: Time resolution ($\sigma(\Delta t)$) as a function of irradiated position for constant level discrimination 
at thresholds of 50 mV (squares) and 250 mV (triangles). 
Right: Time resolution ($\sigma(\Delta t)$) as a function of irradiated position for times calculated at constant fraction of the amplitude 
for fractions of 10\% (squares) and 50\% (triangles). 
\label{thrAndfrac}
}
\end{figure}

Time resolution is determined based on the distribution of time differences measured at a fixed point of irradiation. 
As an example, Fig.~\ref{twoOnOne} presents time difference distributions ($\Delta t = t_R - t_L$) measured by irradiating the scintillator 
strip close to the left photomultiplier (x~=~1.2~cm, right maximum) and at the position close to the right photomultiplier 
(x~=~28.8~cm, left maximum). The times of pulses on both sides $t_L$ and $t_R$ were determined calculating 
the time when the signal crosses a given threshold voltage, as it is illustrated in Fig.~\ref{shapes}.  
The left panel of Fig.~\ref{thrAndfrac} presents the resolution of the time difference measurement as a function of 
the irradiation position.
As expected, due to the “time walk” effect the resolution determined when applying threshold at 250~mV is worse with respect 
to the one obtained at 50~mV. A value of 50~mV was chosen as a 2.5~$\sigma$ 
of a typical 
electronic
noise level 
equal to 20~mV~($\sigma$). 
Right panel of Fig.~\ref{thrAndfrac} shows results obtained when determining the time at the constant fraction of the amplitude.  
From Fig.~\ref{thrAndfrac} one can infer that the time resolution 
is fairly independent of the irradiation position if time is determined 
for constant fraction of the amplitude  as well as for low threshold 
at a constant level (50~mV). It is also visible that in the case of 
the larger threshold (250~mV) resolutions become significantly worse at the edges of the scintillator strip which again is due to the 
“time walk” effect. The resolution of the time difference ($\Delta t = t_R - t_L$) measurement at the center of the scintillator was determined 
to be $\sigma(\Delta t)~=~(153~\pm~2)$~ps both for constant level discrimination 
at 50~mV and constant fraction threshold of 10\% of the amplitude.  
It is important to stress that results shown in Figs.~\ref{twoOnOne} and~\ref{thrAndfrac}
were obtained taking into account only signals with the number of photoelectrons larger than 75.  
This corresponds to the resolution of about 77~ps ($\sigma$) 
for the determination of the interaction moment ($t_{hit}$) which may be expressed
as the average of times measured at the left and right photomultipliers independently of the hit position:
\begin{equation}
	\frac	{(t_R + t_L)}{2} = 	\frac	{\left(t_{hit} + \frac {L-x}{v_{eff}} + t_{hit} + \frac {x}{v_{eff}} \right) } {2} =   t_{hit} + \frac{L}{2 v_{eff}},
\end{equation}
where $v_{eff}$ denotes the effective velocity of the light signal inside the scintillator, 
and  L and x were defined in Fig.~\ref{uklad}. The constant time delays of electronics 
were omitted in the above equation for the sake of simplicity.  
Thus the uncertainty of the measurement of  $t_{hit}$ may be expressed as:
\begin{equation}
	\sigma (t_{hit}) = \frac {\sqrt {  \sigma (t_L) ^2 + \sigma (t_R) ^ 2 }} {2} = \frac{\sigma (\Delta t)}{2}.
\end{equation}
\section{Spatial resolution}
\begin{figure}[h!]
       \centerline{\includegraphics[width=0.49\textwidth]{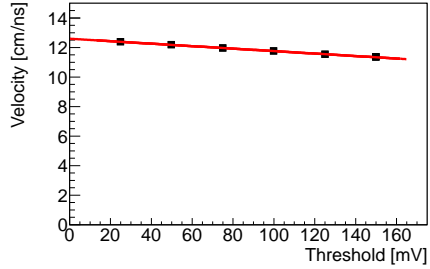}}
\caption{ 
Effective speed of light inside scintillator strip as a function of the applied threshold. 
Superimposed line represents result of the fit of the second order polynomial to the data.
\label{lineAndCompar}
}
\end{figure}

In the first approximation, a hit position along the scintillator strip may be determined based 
on the time difference of light signals arrival to the left and right photomultipliers using 
the following formula:
\begin{equation}
	x = \frac{\Delta t \cdot  v_{eff}}{2},
\end{equation}
which may be derived from the relation:
\begin{equation}
	\Delta t = (t_R - t_L) = t_{hit} + \frac{L-x}{v_{eff}} - t_{hit} - \frac{x}{v_{eff}} = \frac {-2x}{v_{eff}} + C. 
\end{equation}
Thus the spatial resolution reads:
\begin{equation}
	\sigma(x) = \sigma(\Delta t)  \frac{v_{eff}}{2}.
        \label{spatial}
\end{equation}
The effective speed of light signals along a scintillator strip ($v_{eff}$) 
is smaller than the speed of light in a scintillator medium because most of photons do not travel 
to the photomultipliers directly but rather undergoes many internal reflections. 
In order to determine the effective speed of light signals in the
tested scintillator the time difference $\Delta t$ was determined 
as a function of the irradiation position x, and $v_{eff}$ was extracted by fitting a straight line to the experimental points.
The determined value of $v_{eff}$ is shown in Fig.~\ref{lineAndCompar}
as a function of the applied threshold. 
The change of $v_{eff}$ with threshold is due to the “walk effect” 
and the variation of the average amplitude of signals as a function of the distance 
between the interaction point and photomultipliers. 
In order to suppress the bias of the determined $v_{eff}$  due to the value of the applied threshold, 
the effective speed of light signals was determined by 
fitting the second order polynomial to the data points and
extrapolating the fitted curve to the threshold of 0 mV, as shown in Fig.~\ref{lineAndCompar}. 
The systematic uncertainty due to the extrapolation method was estimated as a difference in results between
the fit with second and first order polynomials, and it was found to be negligible. 
The resulting effective speed of light is equal to $v_{eff}~=~(12.61~\pm~0.05_{stat}~\pm~0.01_{sys})$~$\frac{cm}{ns}$.
The determined velocity is in the range of values obtained so far for signals in the plastic 
scintillator bars (see e.g.~\cite{charpak,kouznetsov,shikaze}).

For the estimation of the position resolution 
we apply in equation~\ref{spatial} the value of $v_{eff}$~=~12.2~cm/ns 
and the value of $\sigma(\Delta t)$~=~153~ps both determined for the threshold of 50~mV.
As a result a spatial resolution of $\sigma(x)~=~0.93~cm$  is established 
for the determination of the interaction point of the annihilation quanta along the strip.

\section{Conclusions}
Properties of a single plastic scintillator module of the J-PET detector were investigated in view of the detection of annihilation 
gamma quanta with energy of 511 keV. The module was built out of BC-420 scintillator strip with dimensions 
of 5~mm x 19~mm x 300~mm which was read out at both sides by Hamamatsu R5320 photomultipliers. 
The measurements were performed using a collimated beam of annihilation quanta 
from the $^{68}$Ge isotope and Serial Data Analyzer 
for sampling of photomultiplier’s signals with 50~ps intervals. 
The determined energy resolution amounts to $\sigma(E)/E \approx 0.044 / \sqrt{E(MeV)}$. 
For the energy deposition ranging from 0.18~MeV to 0.34~MeV
the established time resolution 
is equal to about 80~ps~($\sigma$) and the hit position resolution 
along the scintillator strip equals to 0.93~cm~($\sigma$). 
The achieved results are promising and as a next step the test will be conducted with 
a dedicated front-end electronics which will allow 
to sample the signals in the domain of voltage (using multi-threshold sampling) with 
the electronic time resolution below 20~ps~\cite{palka}.

\section{Acknowledgements}
We acknowledge technical and administrative support of T.~Gucwa-Ry\'s, A.~Heczko, M.~Kajetanowicz, G.~Konopka-Cupia\l, 
J.~Majewski, W.~Migda\l, A.~Misiak, and the financial support from the Polish National Center for Development 
and Research through grant INNOTECH-K1/IN1/64/159174/NCBR/12, 
the Foundation for Polish Science through MPD programme, 
the EU and MSHE Grant No. POIG.02.03.00-161 00-013/09, 
Doctus - the Lesser Poland PhD Scholarship Fund, 
and Marian Smoluchowski Krak\'ow Research Consortium "Matter-Energy-Future".

\end{document}